\documentclass[12pt]{iopart}
\usepackage{iopams}

\newtheorem{theorem}{Theorem}[section]

\newtheorem{remark}[theorem]{Remark}
\newlength{\vshift}
\newlength{\hshift}
\usepackage{amssymb,amsopn}

\begin{document}
\begin{flushright}
\end{flushright}
\title{On a polynomial transformation of hypergeometric equations,  Heun's differential equation and exceptional Jacobi polynomials}
%
%
%
%
%
\author{Mahouton Norbert Hounkonnou$ ^1$  and Andr\'e Ronveaux $ ^2$ }
\address{$^1$International Chair of Mathematical Physics
and Applications (ICMPA-UNESCO Chair), University of
Abomey-Calavi, 072 B. P.: 50 Cotonou, Republic of Benin\\
$ ^2$D\'epartement de Math\'ematiques,
2, Chemin du Cyclotron, B-1348, Louvain-la-Neuve, Belgium
}
\eads{\mailto{norbert.hounkonnou@cipma.uac.bj},
\mailto{andre.ronveaux@uclouvain.be}}
%
\begin{center}
\today
\end{center}
\begin{abstract}
This paper addresses a general method of polynomial transformation of hypergeometric equations.
Examples of  some classical special  equations of mathematical physics are generated.  Heun's equation and exceptional Jacobi polynomials are also treated.
\end{abstract}
\noindent
 {\bf AMS Subject Classification:} 35-02; 35A25; 35C10.

\vspace{.08in}
 \noindent \textbf{Keywords}: Fuchsian equation, hypergeometric differential equations, Heun's equations, 
singular points, Jacobi, Laguerre and Hermite equations,  exceptional polynomials.\\
%

\section{Introduction}
The singularities of linear ordinary differential equation (ODE), Fuchsian or not Fuchsian, can move,
change, coalesce, etc.  by a transformation acting on the independent
variable $x$, or on the dependent variable $y$ in the equation
$L\{y(x)\} = 0.$ For instance, in mathematical physics, one uses to transform a
given equation, like the Schroedinger equation of the harmonic oscillator,
 into a well known equation, as the Hermite differential
equation, changing both the independent variable (rescaling) and the
dependent variable (wave function). In this exemple,  the alone
singularity at the infinity stays at infinity. 

%
%
 In $1971,$  Kimura  \cite{r1} investigated in detail all Fuchsian differential equations $F\{y(x)\}=0$ reducible to the hypergeometric equation $H\{z(x)\}=0$ by a linear transformation $y(x) =P_0(x)z(x) + P_1(x)z^\prime (x)$,with
$P_0(x)$ and $P_1(x)$  rational functions of $ x$. In order to compute $P_0(x)$ and
$P_1(x)$, he  assumed the same set of singularities and the same
monodromy group for $H\{y(x)\}=0$ and $F\{z(x)\}=0$, giving information about the
structures and properties of $P_0(x)$ and$P_1(x)$.
The equation $ F\{y(x)\}= y''+ A_1(x)y' + A_2(x)y= 0,$  $( y=y(x))$,
being Fuchsian and written as a generalized Heun's equation contains
regular singular points with many parameters, but not all free, in order to
eliminate logarithmic situations and irreductibility (equation not
factorizable). Many theorems and properties allow to compute heavily $P_0(x)$
and $P_1(x)$ for each peculiar situations, but without using the coupled
differential equations satisfied by $P_0(x$) and $P_1(x)$  given  by the
author.
(See \cite{r2}- \cite{r13} for more details on Heun's equations and orthogonal polynomials).

The aim of this paper is to reverse this approach, allowing to eliminate the
Fuchsian constraints: inject an {\it arbitrary} linear transformation $y=A(x)z +B(x)z'$
in the hypergeometric equation and build  a second order differential
equation for $F(z(x))$, not always Fuchsian, but with $2$  families of arbitrary
differentiable functions, polynomial or not.
This approach allows also to use any solution $z(x)$ of the hypergeometric
equation, polynomial or not,  including possibly solutions of second kind.
It is surprising that  recently introduced new orthogonal polynomials, also called  exceptional polynomials
(Jacobi, Laguerre, etc.) \cite{r6, r9}, can be also  built  from the same linear transformation long time ago given in 
 the above mentioned seminal paper by Kimura. Recall that the concept of exceptional orthogonal polynomials was introduced by G\'omez-Ullate  {\it et al} \cite{r14, r15}. Within the Sturm-Liouville theory they constructed $X_1$ Laguerre and $X_1$ Jacobi polynomials, which turned out to be the first members of the infinite families. Then Quesne and collaborators \cite{r16, r17} reformulated their results in the framework of quantum mechanics and shape-invariant potentials. The merit of quantum mechanical reformulation resides in the fact that the orthogonality and completeness of the obtained eigenfunctions are guaranteed. Besides, the well established solution mechanism of shape invariance combined with the Crum's method \cite{r18}, or the so-called factorization method \cite{r19}, or the susy quantum mechanics  \cite{r20} is available. A nice discussion on these aspects is presented in \cite{r21} and references therein.

%
%

\section{General setting}
Consider the hypergeometric equation:
\begin{eqnarray}\label{eq1}
 \sigma(x)z^{\prime\prime}(x) +\tau(x)z^{\prime}(x) +\lambda z(x)= 0,
\end{eqnarray}
where $  \sigma\equiv\sigma(x)$  is a polynomial of degree less or equal to $2,$ and  $\tau\equiv\tau(x)$ a polynomial of degree exactly equal to $1;$
$\lambda$ is a constant. Let the following transformation:
\begin{eqnarray}\label{eq2}
 y(x)= A(x)z(x) + B(x)z^{\prime}(x)
\end{eqnarray}
with $A= A(x)$ and $B= B(x)$ polynomials of degrees $r$ and $s$, respectively.
This transformation appears for instance in the recent development of exceptional $X_l$ Laguerre and Jacobi polynomials \cite{r6, r9} and also
in the problem of reducing Fuchsian ordinary differential equations into hypergeometric equations (ODEs) like Heun's equations \cite{r11, r12, r13}.

In this work, we build a general linear second order ordinary differential equation
 ${\cal L}_2[y(x)]= 0$ satisfied by the function
$y(x)$, polynomial or not, and we investigate a set of situations.

The first derivative of (\ref{eq2}) gives, using (\ref{eq1}):
\begin{eqnarray}\label{eq3}
  \sigma y^{\prime}= {\bar A}z + {\bar B}z^{\prime}
\end{eqnarray}
with ${\bar A}= \sigma A^\prime - B\lambda,$  ${\bar B}= \sigma A + \sigma B^\prime - \tau B.$

The derivative of (\ref{eq3}) can be written as
\begin{eqnarray}\label{eq4}
  \sigma(\sigma y^{\prime})^\prime= {\bar{\bar A}}z + {\bar{\bar B}}z^{\prime}
\end{eqnarray}
with
${\bar{\bar A}}= \sigma \bar A^\prime - \bar B\lambda,$  $\bar {\bar B}= \sigma \bar A + \sigma \bar B^\prime - \tau \bar B$
which can be expanded in terms of $\sigma, \;\;\tau, \;\;\lambda$ as follows:
\begin{eqnarray}
{\bar{\bar A}}&=&A^{\prime\prime}\sigma^2+\sigma
A^\prime\sigma^\prime-2\lambda\sigma B^\prime-\lambda
A\sigma+\lambda B\tau\\
 {\bar{\bar B}}&=&2A^\prime\sigma^2-\lambda\sigma B+\sigma
 A\sigma^\prime+\sigma^2 B^{\prime\prime}+\sigma
 B^\prime\sigma^\prime\nonumber\\
 &-& 2 \sigma B^\prime \tau-\sigma B \tau^\prime-\tau A\sigma+B
 \tau^2.
\end{eqnarray}
We then arrive at the following determinantal equation giving the ordinary differential equation satisfied by $y(x):$
\begin{equation}\label{eq5}
{\cal L}_2[y(x)]=\left|
\begin{array}{ ccc}
       y & A & B\\
\sigma y^\prime & \bar A & \bar B\\
\sigma (\sigma y^\prime)^\prime & \bar{\bar A} & \bar{\bar B}
       \end{array}
\right|
=0.
\end{equation}
The following presentation is also useful:
\begin{equation}\label{eq5b}
{\hat{\cal L}}_2[y(x)]=\left|
\begin{array}{ ccc}
       y & A & B\\
\sigma y^\prime & \bar A & \bar B\\
\sigma ^2 y^{\prime\prime }& {\hat {A}} & {\hat{ B}}
       \end{array}
\right|
=0
\end{equation}
with 

\begin{eqnarray}
{\hat{ A}}&=&\sigma {\bar A}^\prime-{\bar B}\lambda-\sigma^\prime{\bar A}
\\
 {\hat{ B}}&=&\sigma {\bar A} + \sigma {\bar B}^\prime- {\bar B}(\tau +\sigma^\prime).
\end{eqnarray}

\begin{remark}The following observations are in order:
 \begin{enumerate} 

  \item The coefficient of $y^{\prime\prime}$ is $\sigma^2(A\bar B - B\bar A)$ and introduces new singular points
depending on the degrees $r$ and $s$ of the polynomials $A$ and $B$.
In general the obtained ODE is no more Fuchsian.
\item This general formulation does not take into account the possible logarithmic solutions and apparent singularities
for ${\cal L}_2[y(x)]= 0$ which strongly restrict the choice of the
polynomials $A$ and $B$. Such a freedom allows to cover a larger
class of equations with $r+s+2$ parameters such as Heun, generalized
Heun equations and their various confluencies and to propose some
solutions.
\item An obvious way to remain in the Fuchsian's class is to choose  $B(x)= \sigma(x)$ or more generally $B(x)= S(x) \sigma(x)$.  Such a simplified choice has been already considered in the work by Kimura \cite{r1} and in recent literature on exceptional
polynomials (see \cite{r6} and references therein).
 \end{enumerate}
\end{remark}
\subsection{Basic choice: $A(x)$ is arbitrary and $B(x)= \sigma(x)$}
The choice $B(x)= \sigma(x)$ generates a simpler equation, eliminating the factor $\sigma^2$ in the coefficient of $y^{\prime\prime}$:

\begin{equation}\label{eq6}
{\cal L}_2[y(x)]=\left|
\begin{array}{ ccc}
       y & A & \sigma\\
 y^\prime & A^\prime - \lambda & A +\sigma^\prime - \tau\\
(\sigma y^{\prime})^\prime & C & D

       \end{array}
\right| =0,
\end{equation}
where
\begin{eqnarray}
C&=&A^{\prime\prime}\sigma+ A^{\prime}\sigma^\prime+\lambda(\tau -2\sigma^\prime -A)\\
D
&=& \sigma(\sigma^{\prime\prime}-\lambda-\tau^\prime+2A^\prime)+
\sigma^\prime(\sigma^\prime -2\tau+A) +\tau (\tau-A)
\end{eqnarray}
yielding the second order differential equation:
\begin{eqnarray}\label{eq7}
{\cal L}_2[y(x)]= \sigma P y^{\prime\prime} + Q y^\prime + R y= 0
\end{eqnarray}
with
\begin{eqnarray}
 P&\equiv& P(x)= A^2 + A(\sigma^\prime - \tau) + \sigma (\lambda -
 A^\prime)\\
Q&\equiv&
Q(x)=A^{\prime\prime}\sigma^2+\sigma(A\tau^\prime+\tau\lambda-\lambda\sigma^\prime-A\sigma^{\prime\prime}-2AA^{\prime})\cr
&+&A\tau(\sigma^\prime+A-\tau)
\\
R&\equiv& R(x)=\sigma[ A^{\prime\prime}(\tau-
A-{\sigma^\prime})+A^\prime(2A^\prime+
\sigma^{\prime\prime}-3\lambda -\tau^\prime) \cr
&+&\lambda(\lambda+
\tau^{\prime}-\sigma^{\prime\prime})]\nonumber\\
&+&\sigma^\prime[2\lambda A +\lambda\sigma^\prime-\lambda
 \tau-\tau A^\prime] +(A-\tau)(\lambda  A
-A^\prime \tau )
\end{eqnarray}

It is worth noticing that appropriate choices of coefficients in the polynomial $A(x)$ reducing equation (\ref{eq6})
to hypergeometric equations
generate known or unknown contiguous relations between the solutions $y(x)$. To cite a few, for instance with
 $\sigma(x)= B(x)= 1 - x^2,$
and $\tau(x)= \beta - \alpha - (\alpha + \beta + 1)x$ and $\lambda=
(n-1)(n + \alpha + \beta),$ the functions $y(x)$ are reduced to the
Jacobi polynomials $P_{n-1}^{(\alpha, \beta)}(x)$ as polynomial solution. 
The equation (\ref{eq7}) coincides, after simplification, with the equation
\begin{eqnarray}
{\cal L}_2[y(x)]&=& ( 1 - x^2)y^{\prime\prime}(x) + [\beta - \alpha - (\alpha + \beta + 2)x]y^\prime(x) \cr 
&+& (n-1)(n + \alpha + \beta)y(x)= 0
\end{eqnarray}
with $y(x)= P_{n-1}^{(\alpha, \beta)}(x)$ as solution. This proves therefore the well known contiguous relation:
\begin{eqnarray}\label{ce}
 (2n+\alpha + \beta)(1-x^2){d\over {dx}} P_{n}^{(\alpha, \beta)}(x)&-& n[\alpha -
 \beta - (2n+\alpha + \beta)x]  P_{n}^{(\alpha, \beta)}(x)\cr
&=& 2(n+\alpha)(n +   \beta)P_{n-1}^{(\alpha, \beta)}(x)
\end{eqnarray}
providing the polynomial coefficient $A(x)$ of $P_{n}^{(\alpha, \beta)}(x)$ after division of the equation (\ref{ce}) by the factor $2(n+\alpha)(n +   \beta).$
 Of course if $\lambda$ is arbitrary,  non polynomial solutions  as
the second kind solutions of (\ref{eq1}) generate similar contiguous relations for the $z(x)$  as for the polynomial solutions $y(x)$.

\subsection{Second choice: $A$ and $B$ are constants}
This trivial case leads to another simpler equation:
\begin{eqnarray}\label{eq8}
{\cal L}_2[y(x)]&=&\left|
\begin{array}{ ccc}
       y & A & B\\
\sigma y^\prime & -\lambda B & \sigma A -  \tau B\\
\sigma (\sigma y^\prime)^\prime & -\lambda(\sigma A -  \tau B) &
\sigma(A\sigma^\prime-B\tau^\prime-A\tau-\lambda B)+B\tau^2
       \end{array}
\right|\cr
&=&0
\end{eqnarray}
providing the second order differential equation:
\begin{eqnarray}\label{eq9}
{\cal L}_2[y(x)]= \sigma P y^{\prime\prime} + Q y^\prime + R y= 0
\end{eqnarray}
with
\begin{eqnarray}
 P&\equiv& P(x)= \sigma(x)[A(A\sigma - \tau B) + \lambda B^2]\\
Q&\equiv& Q(x)=\sigma^2(\tau A^2+ A B \tau^\prime)+\sigma (\lambda
B^2\tau+\lambda\sigma^{\prime} B^2 -A B \tau^2-\sigma^{\prime}A
B\tau)\\
R&\equiv& R(x)=\lambda A^2 \sigma^2+\sigma(\lambda^2B^2 -\lambda A B
\sigma^{\prime} +\lambda B^2\tau^\prime-\lambda B A \tau ).
\end{eqnarray}
This equation is also not hypergeometric in general, but using suitable choices we can deduce again contiguous relations.
\begin{itemize}
 \item
As illustration, if $\sigma(x)= x,$ $\tau(x)= 1 + \alpha - x$ and
$\lambda= n,$ the $z(x)$ are Laguerre polynomials
$L_n^{(\alpha)}(x).$ The choice $A= 1= -B$ gives again a Laguerre
polynomial for $y(x)$ confirming the well known relation
\begin{equation}
L_n^{(\alpha)}(x) - [L_n^{(\alpha)}(x)]^\prime= [L_{n+1}^{(\alpha)}(x)]^\prime= L_n^{(\alpha+1)}(x).
\end{equation}
\item If $\sigma= 1$ and $\tau= -2x,$ the equation (\ref{eq7}) takes the form:
\begin{eqnarray}
 {\cal L}_2[y(x)]&=&y^{\prime\prime}(A^2+B^2\lambda+2ABx) \cr
&-& y^{\prime}[2AB+2x(A^2+B^2\lambda)+4x^2AB]\cr
& + &y\lambda[B^2(\lambda-2)+A(A+2xB)]= 0
\end{eqnarray}
which can be further simplified to give, for the particular case of $B= 0$ with arbitrary $A$, the well known Hermite equation:
\begin{eqnarray}
{\cal L}_2[y(x)]= y^{\prime\prime} - 2xy^{\prime} + \lambda y= 0.
\end{eqnarray}
In the opposite, when $A= 0$ and $B$ is an  arbitrary function, the equation (\ref{eq9}) is transformed into a modified Hermite equation
\begin{eqnarray}
{\cal L}_2[y(x)]= y^{\prime\prime} - 2xy^{\prime} + {\bar \lambda} y= 0\,\,\mbox{with}\,\,.{\bar \lambda}= \lambda-2.
\end{eqnarray}
\end{itemize}

\section{Link with the Heun equations and exceptional Jacobi polynomials}

With $A=\alpha x+\beta$ and $\sigma=x^2-x$, the degrees of $P, Q, R$ in equation (\ref{eq7})  are
$2, 3$  and $2,$ respectively. In order to generate a HEUN equation 

\begin{equation}
y^{\prime\prime} + \left[ {\gamma \over x} + {\delta \over {x-1}}+
{\epsilon\over{x-\mu}}\right]y^\prime +{{\alpha\beta x + \rho}\over
{x(x-1)(x-\mu)}}y= 0,
\end{equation}

the coefficients $\alpha,\beta$ and  $\tau$ must be chosen such that $P(x)$ reduces to $x-\mu$,
and $Q(x)=(x-\mu)Q_1(x)$ and $R(x)=(x-\mu) R_1(x),$  where   $Q_1(x)$ and $R_1(x)$ are polynomials. 
Other appropriate choices also produce confluent Heun equations \cite{r11}.
All  these situations generate Heun equations with explicit solutions
$y(x) =A(x)z(x) + B(x) z^\prime (x).$
Kimura gives in \cite{r1}  solutions in  two relevant cases: $(i)$ for $\epsilon = -1$ with $A(x)$ of degree $1$ and $B(x)
=x(x-1)$, and $(ii)$  for $\epsilon = -2$ with $A(x)$ of degree $2$ and $B(x)$
of degree $3$. The  Kimura  method  is very nice, but 
 complicated and 
restrictive, the confluent equations being excluded. Its intrinsic
complexity 
resides in the step $\epsilon= -1$ to $- 2$ increasing the degree
of polynomials $A(x)$ and $B(x)$. In our approach,  the problem is entirely algebraic, even not excluding,  of
course,  also some difficulty.

Finally, let us mention that the exceptional $X_1-$Jacobi polynomials investigated in  \cite{r6} can be also  easily retrieved from our method. Indeed,   
set, for $g, h \notin \{ -1/2, -3/2, -5/2, \ldots \}:$
$$\zeta(\eta)={{g-h}\over 2}\eta + {{g+h+1}\over 2},{\tilde{\zeta}}(\eta)={{g-h}\over 2}\eta + {{g+h+3}\over 2}.$$ and  consider
 the polynomials
$$A(\eta)= {{1}\over{k+h+{{1}\over {2}}}}\Big(h+{1\over 2}\Big){\tilde{\zeta}}(\eta)$$
and
$$B(\eta)= {{1}\over{k+h+{{1}\over {2}}}}\Big(1+\eta\Big)\zeta(\eta)$$
of degrees $1$ and $2,$ respectively. Let
 $z(\eta)$ be the Jacobi polynomial parametrized as \cite{r6}:
\begin{eqnarray}
P_k(\eta)= {{(g+{1 \over 2})_k}\over {k!}}\sum_{j=0}^k {{(-k)_j(k+g+h+2)_j}\over{j!(g+{1\over 2})_j}}\Big({{1-\eta}\over 2}\Big)^j.
\end{eqnarray}
Then the  transformation (\ref{eq2}) gives the  $X_1-$ Jacobi polynomials $y(\eta)\equiv {\hat {P_k}}(\eta)$ satisfying the following differential equation:
\begin{eqnarray}
&&(1-\eta^2)y^{\prime\prime} (\eta)+ \Big(h-g-(g+h+3)\eta -2{{(1-\eta^2)\zeta^\prime(\eta)}\over{\zeta(\eta)}}\Big)y^\prime(\eta)\cr
&+& \Big(-{{2(h+{1\over 2})(1-\eta){\tilde{\zeta}^\prime(\eta)}}\over{\zeta(\eta)}}+k(k+g+h+2)+g-h\Big)y(\eta)=0.
\end{eqnarray}
Transforming $\eta$ into a new variable and assigning adequate relations between parameters lead to specific Heun's differential equations as shown, for instance in \cite{r6}, with the particular variable change $\eta= 1-2x.$ For more details about Heun's equation and differential equation describing the exceptional Jacobi polynomials, see \cite{r6} and references therein.

\section*{Acknowledgments}

The authors thank  Dr Alain Moussiaux for testing some algebraic equations with CONVODE software.


\end{document}